\documentclass[12pt]{article}
\usepackage{color}
\usepackage{amsmath,amssymb}
\usepackage{amscd}
\usepackage[dvipdfmx]{graphicx} 
\usepackage{epstopdf}
\usepackage{latexsym}

\allowdisplaybreaks[1]

\newcommand{\eq}[1]{(\ref{#1})}
\newcommand{\nn}{\nonumber}
\newcommand{\ds}{\displaystyle}

\newcommand{\vev}[1]{\left\langle #1 \right\rangle}

\renewcommand{\thefootnote}{\fnsymbol{footnote}}

\makeatletter
\@addtoreset{equation}{section}

\begin{document}


\begin{titlepage}
\thispagestyle{empty} 

\begin{flushright}
March 18, 2019
\end{flushright}

\vspace{1.0cm}
%

%
\vspace{1.2cm}

\begin{center}
\noindent{\Large \textbf{
Quantum Annealing of Vehicle Routing Problem \\
\vspace{0.2cm}
with Time, State and Capacity%
\footnote{This is a pre-print of an article to be published in Feld S., Linnhoff-Popien C. (eds)
,
Quantum Technology and Optimization Problems (QTOP 2019), Lecture Notes in
Computer Science, vol 11413, Springer, 2019. The final authenticated version is available online at DOI: {\tt http://doi.org/10.1007/978-3-030-14082-3}
}
}}
\end{center}

\vspace{1cm}

\begin{center}
\noindent{Hirotaka Irie,\footnote{hirotaka\_irie@denso.co.jp}$^{,a}$ 
Goragot Wongpaisarnsin,\footnote{goragot@th.nexty-ele.com}$^{,b}$ 
Masayoshi Terabe,\footnote{masayoshi\_i\_terabe@denso.co.jp}$^{,a}$ 
Akira Miki\footnote{akira\_miki@denso.co.jp}$^{,a}$ 
and Shinichirou Taguchi\footnote{shinichrou\_taguchi@denso.co.jp}$^{,a}$ }
\end{center}

\vspace{0.3cm}

\begin{center}
{\em
 Information Electronics R\&I Department, Electronics R\&I Division, \\
Advanced Research and Innovation Center, \\
DENSO Corporation, Tokyo Office, Tokyo 103-6015, Japan$^{a}$ 
 \\
 Contents Development and Distribution Department, \\ 
Toyota Tsusho Nexty Electronics Co., LTD., 
15th-16th Floor, Mercury Tower, 540 Ploenchit Road,  Lumpini, Pathumwan, Bangkok 10330, Thailand$^{b}$ 
 \\
}
\end{center}

\vspace{0.3cm}

\begin{abstract}
We propose a brand-new formulation of capacitated vehicle routing problem (CVRP) as quadratic unconstrained binary optimization (QUBO). The formulated CVRP is equipped with {\em time-table} which describes time-evolution of each vehicle. Therefore, various constraints associated with time are successfully realized. With a similar method, constraints of {\em capacities} are also introduced, where capacitated quantities are allowed to increase and decrease according to the cities which vehicles arrive. As a bonus of capacity-qubits, one also obtains a description of {\em state}, which allows us to set a variety of traveling rules, depending on each state of vehicles. As a consistency check, the proposed QUBO formulation is also evaluated by quantum annealing with D-Wave 2000Q. 
\end{abstract}

\end{titlepage}

\newpage

\renewcommand{\thefootnote}{\arabic{footnote}}
\setcounter{footnote}{0}


\tableofcontents



\section{Introduction}

Vehicle Routing Problem (VRP) \cite{VRP-first-reference} is a basic mathematical problem related to optimization of planning, logistics, and transportation. Owing to the recent interests in quantum annealing machines, which were first studied theoretically by Kadowaki-Nishimori \cite{Kadowaki-Nishimori1998} and made available commercially by D-Wave Systems Inc.~\cite{DWaveCite}, the investigation of VRP as a quadratic unconstrained binary optimization (QUBO) has become very important, particularly in an attempt to achieve quantum-mechanical optimization of real-world problems encountered in our daily life concerning various mobility services. 

VRP is a generalization of Traveling Salesman Problem (TSP), i.e.,~a problem to find the traveling path that has the lowest cost. Similarly, the purpose of VRP is to find the best routing scheme for multiple vehicles that achieves the lowest cost under various circumstances; for instance, each city should be visited exactly once. QUBO formulation of the TSP is found in \cite{Lucas2014}; it was constructed by adding the square of linear-constraint functions to the associated cost functions \cite{HopfieldTank1985, HopfieldTank1986, MezardParisi1985} using Lagrange-like multipliers. 
A straightforward extension of this QUBO formulation can also be applied to VRP by introducing several copies of QUBO systems for TSP. 
However, such formulation intrinsically suffers from the strict concept of {\em time}. 
It does have a concept of time-step, but it is generally not equivalent to the concept of time in many practical applications. 
Obviously, if one of the cities is located far from the rest, then traveling there will take more time than traveling to other cities. 
In such case, the conventional time-step formulation cannot describe the time, which flows commonly and homogeneously  for all vehicles. 
Therefore, the introduction of time in the conventional QUBO formulations is the main obstacle in formulating various important VRP constraints associated with time such as time-window, (non-)simultaneous arrivals, and chronological variation of cost.

In addition to the concept of time, it is also important to introduce the concept of 
{\em capacity} (i.e.,~capacitated VRP or CVRP), which may describe the capacity of carrying passengers or packages. 
There are some attempts toward a QUBO formulation of CVRP: 
one tackles a possibility of constraint terms describing inequality \cite{BrainPad2017}; 
and another utilizes a hybrid cluster algorithm combined with the TSP QUBO systems \cite{QASAR2018}. 
In this paper, we propose a different approach to investigate a concise QUBO formulation. 
In fact, capacity and time are similar as
the description of time already implies that vehicles should travel within their own capacity of time. 
As further explained, these two concepts are similarly implemented by introducing a new kind of interactions that depend on the cities of departure and destination only. 
Furthermore, by introducing {\em capacity-qubits} (See Section \ref{section-capacity-qubits}), 
one can realize multiple capacitated variables which can be allowed to increase and decrease (i.e.~pickup and delivery) during each travel. 

However, there are further applications of this approach. 
As a bonus of capacity-qubits, one can also describe the concept of {\em state}. In particular, cost and time-duration can change depending on each state of vehicles. As a simplest example, we demonstrate a two-state model: arrival-state and departure-state. The mean time from arrival to departure is the duration of visits. We can also set up how long each vehicle stays depending on the cities visited by vehicles.

The organization of this paper is as follows: The brand-new QUBO formulation of VRP is introduced with time-table (in Section \ref{Section-time}), with capacity-qubits (in Section \ref{section-capacity-qubits}), and with the concept of state (in Section \ref{Section-state}). Other related constraints from real-world applications are in Section \ref{Section-other}, and validity of our models is discussed in Section \ref{section-CC-cond}. Conclusion and discussion are presented in Section \ref{Section-conclusion}. 

\section{Time-table in TSP/VRP \label{Section-time}}

The first new ingredient introduced to the TSP/VRP concept in this paper is {\em time}. In particular, we introduce {\em time-table} in the TSP/VRP formulation. 
First, we would prepare binary qubits parametrized by three integers $(\tau,a,i)$: 
\begin{align}
x_{\tau,a}^{(i)} \qquad  \bigl( 1 \leq \tau \leq T, \quad 1 \leq a \leq N,\quad  1\leq i \leq k \bigr),  \label{eq-VRP-qubits}
\end{align}
where $\tau$ parametrizes each {\em time-interval} of the time-table, with the assumption that there are $N$ cities to visit and $k$ vehicles are present. 
The time-interval means that we divide the total time into several time units as follows: 
\begin{align}
(\tau) \qquad \mapsto \qquad \bigl[t_\tau, t_{\tau+1} \bigr) \qquad \bigl( 1\leq  \tau \leq T-1 \bigr), \qquad t_{\tau +1} - t_\tau \equiv \Delta t. 
\end{align}
Herein $\Delta t$ is the unit of time-division.%
\footnote{Note that the unit of time-division can also depend on time $(\tau)$ and vehicle $(i)$, i.e.,~$\Delta t_\tau^{(i)} \equiv t_{\tau+1}^{(i)} - t_\tau^{(i)}$. 
Although such a generalization is also important for some applications, we keep the uniform value $\Delta t_{\tau}^{(i)} = \Delta t$ for the sake of simplicity. }
For example, three hours from 9:00 AM to 12:00 AM can be divided into nine intervals with twenty-minute duration each. 
Suppose a vehicle $(i)$ does (or does not) arrive at a city $(a)$ in a time-interval $(\tau)$, the binary qubits $x_{\tau,a}^{(i)}$ takes the following value: 
\begin{align}
x_{\tau,a}^{(i)} = 1 \quad \bigl( \text{arriving} \bigr), \qquad x_{\tau,a}^{(i)} = 0 \quad \bigl( \text{not arriving} \bigr). \label{eq-VRP-qubits-arrival}
\end{align}

Since the conventional QUBO formulation of TSP/VRP always assumes one-step forward, 
we would invent a new kind of interaction which describes how much time each vehicle spends for each travel. 
Thus, we would first introduce {\em time-duration matrices} $\bigl( n_{ab}^{(\tau)} \bigr)_{1\leq a \neq b \leq N}$ 
as well as cost matrices $\bigl(d_{ab}^{(\tau)}\bigr)_{1 \leq a \neq b \leq N}$ 
for each time-interval $(\tau)$ as%
\footnote{Here $\lceil x \rceil$ ($x\in \mathbb R$) is the minimum integer which is greater than or equal to $x$, i.e.,~$x \leq \lceil x \rceil < x+1$ and $\lceil x \rceil \in \mathbb Z$.  }
\begin{align}
n_{ab}^{(\tau)} &=  \biggl \lceil \frac{  \bigl(  \text{time spent from a city $(b)$ (at time $\tau$) to a city $(a)$} \bigr)  }{\Delta t} \biggr \rceil \geq  1, \\
d_{ab}^{(\tau)} &= \bigl( \text{cost spent from a city $(b)$ (at time $\tau$) to a city $(a)$} \bigr). 
\end{align}
Thus, our proposed Hamiltonian $\mathcal H$ can then be written as follows: 
\begin{align}
\mathcal H  & = 
\sum_{ 
\left\{
\begin{subarray}{c}
1\leq a \neq b \leq N \cr
1\leq \tau \leq T -1 \cr
1\leq i \leq k
\end{subarray}
\right\}
} \biggl(
\underbrace{ \frac{d_{ab}^{(\tau)} - \mu}{\rho} \times x_{\tau + n_{ab}^{(\tau)}, a}^{(i)} \, x_{\tau,b}^{(i)} }_{(\ast 1)}
+ \sum_{1\leq \delta \tau \leq n_{ab}^{(\tau)} - 1} 
\underbrace{ \lambda \times x_{\tau + \delta \tau, a}^{(i)} \, x_{\tau, b}^{(i)}}_{(\ast 2)} \biggr) +   \nn\\
& +  \underbrace{
 \lambda\times \biggl( 
\sum_{ 
\left\{
\begin{subarray}{c}
1\leq a < b \leq N \cr
1\leq \tau \leq T \cr
1\leq i \leq k 
\end{subarray}
\right\}
} 
x_{\tau,a}^{(i)} \, x_{\tau,b}^{(i)}    +
\sum_{ 
\left\{
\begin{subarray}{c}
1\leq a \leq N \cr
1\leq \tau \neq \tau' \leq T \cr
1\leq i \leq j \leq k 
\end{subarray}
\right\}
} 
x_{\tau',a}^{(i)} \, x_{\tau,a}^{(j)}    +
\sum_{ 
\left\{
\begin{subarray}{c}
1\leq a \leq N \cr
1\leq \tau \leq T \cr
1\leq i <  j \leq k 
\end{subarray}
\right\}
} 
x_{\tau,a}^{(i)} \, x_{\tau,a}^{(j)}
}_{(\ast 3)}
\biggr). 
 \label{Hamiltonian-1}
\end{align}
The new form of interaction is introduced in the first line: $(\ast 1)$ and $(\ast 2)$
\begin{itemize}
\item [1. ] The first term $(\ast 1)$ gives the cost of travel from a city $(b)$ (departed at $\tau$) to a city $(a)$ (the arrival is at $\tau' = \tau + n_{ab}^{(\tau)}$). 
\item [2. ] The second term $(\ast 2)$ forbids {\em any early arrivals} at a city $(a)$ (i.e.,~at $\tau'$ in range $1 \leq \tau' - \tau < n_{ab}^{(\tau)}$). 
This is realized by repulsive interactions set forward from the departure city $(b)$ to the arrival city $(a)$.
\end{itemize}
The other terms $(\ast 3)$ are obtained from the basic constraints of penalty terms: 
\begin{itemize}
\item [1. ] Any vehicle $(i)$ in any time-interval $(\tau)$ will not arrive at two different cities $(a)$ and $(b)$ simultaneously: 
\begin{align}
\lambda \times x_{\tau,a}^{(i)} \, x_{\tau,b}^{(i)} \qquad \bigl( \forall a \neq \forall b, \forall i, \forall \tau \bigr).  \label{Const-Eq-ab}
\end{align}
\item [2. ] If a vehicle $(i)$ arrives at a city $(a)$ in a time-interval $(\tau)$, then that vehicle $(i)$ will not arrive at a city $(a)$ in any other time-interval $(\tau')$: 
\begin{align}
\lambda \times x_{\tau,a}^{(i)} \, x_{\tau', a}^{(i)}\qquad \bigl( \forall a, \forall i, \forall \tau \neq \forall \tau' \bigr). 
\end{align}
\item [3. ] If a vehicle $(i)$ arrives at a city $(a)$ in a time-interval $(\tau)$, then the other vehicles $(j)$ will not arrive at a city $(a)$ in any time-interval $(\tau')$:
\begin{align}
\lambda \times x_{\tau,a}^{(i)} \, x_{\tau', a}^{(j)} \qquad \bigl( \forall a, \forall i \neq \forall j, \forall \tau, \forall \tau' \bigr). 
\end{align}
\end{itemize}
At this point, we no longer employ the conventional method of constraint functions (i.e.,~square of linear-constraint functions, applied in \cite{Lucas2014}). Instead, we have introduced {\em an additional parameter} $\mu\, (>0)$ around the traveling cost $d_{ab}^{(\tau)}$, as well as the standard parameters $\rho\,  (>0)$ and $\lambda \, (>0)$. 
\begin{itemize}
\item Overall negative shift of costs (delivered by $\mu$) replaces the role of the negative linear terms generated in the conventional method of constraint functions (See Section \ref{section-CC-cond}). 
\item Up to the overall scaling, at least two parameters ($\mu, \rho$) should be adjusted properly to optimize the performance of each Ising machine.  The remaining $\lambda$ is adjusted by total scaling to the maximum value suited for each Ising machine. 
\item However, as discussed in Section \ref{section-CC-cond}, the standard value of the parameters $\mu$ and $\rho$ are selected as follows: 
\begin{align}
\mu = d_{\rm max},\qquad \rho = \frac{d_{\rm max} - d_{\rm min}}{\lambda}.  \label{Eq-mu-rho-parameter-choice}
\end{align}
\end{itemize}
Generally, the initial starting points of the vehicles can be implemented using boundary condition of the binary qubits. For instance, one can choose 
\begin{align}
x_{1,a}^{(i)} = \delta_{a,s_i} & \qquad \bigl( \text{the starting point of vehicle $(i)$ is a city $(s_i)$} \bigr). 
\end{align}
Conversely, to set the final destination, we turn off several binary qubits from which one cannot reach the final destination before exceeding the upper time limit $T$ 
(where the final destination of vehicle $(i)$ is the city $(e_i)$): 
\begin{align}
x_{\tau,a}^{(i)} = 0  \qquad \text{when \quad $\tau + n_{e_i, a}^{(\tau)} > T$},
\end{align}
Additionally, the forward interactions from the final destination should be replaced as follows:
\begin{align}
&\biggl( \frac{d_{ab}^{(\tau)} - \mu}{\rho} \times x_{\tau + n_{ab}^{(\tau)}, a}^{(i)} \, x_{\tau,b}^{(i)} 
+ \sum_{1\leq \delta \tau \leq n_{ab}^{(\tau)} - 1} \lambda \times x_{\tau + \delta \tau, a}^{(i)} \, x_{\tau, b}^{(i)} \biggr)\bigg|_{b = e_i}  \nn\\
&\quad \underset{\text{replace}}{\longrightarrow}  \quad \biggl( \sum_{1\leq \delta \tau \leq T- \tau} \lambda \times x_{\tau + \delta \tau,a}^{(i)} x_{\tau, e_i}^{(i)} \biggr), 
\end{align}
to make sure that any travel after arriving at the final destination is forbidden. 

As this TSP/VRP formulation can deal with time-scheduling, we refer it to as {\em time-scheduled TSP/VRP} or {\em TS-TSP/VRP}.

\section{Multiple-capacitated TSP/VRP \label{section-capacity-qubits}}

In the last section, we have introduced the concept of time and, as is mentioned in Introduction, time is a kind of capacitated variable whose consumption is accumulated until full of its ``capacity''. Therefore, one can replace the concept of time by the concept of capacity for ``a monotonically increasing/decreasing capacitated variable''. This replacement is also useful in some practical applications, especially in a case of saving the number of qubits. However, such a concept of time/capacity only resolves the problem of single-capacitated TSP/VRP. 

\begin{figure}
\includegraphics[width=\textwidth]{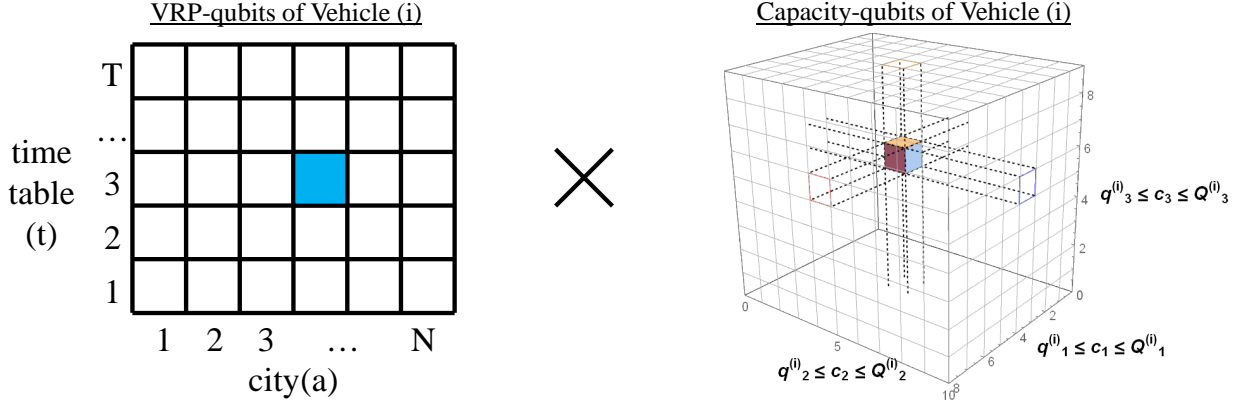}
\caption{Binary qubits representation as boxes. } \label{capacity-qubit-figure}
\end{figure}

In this section, we shall introduce the concept of multiple capacities in addition to our time-scheduled TSP/VRP formulation, which is referred to as {\em time-scheduled multiple-capacitated VRP} or {\em TS-mCVRP}. 
Addition of multiple-capacity can be achieved by adding {\em capacitated variables} $(c_1,c_2,\cdots,c_M)$ in the binary qubits of Eq.~\eq{eq-VRP-qubits} as follows:
\begin{align}
x_{\tau,a|c_1,c_2,\cdots,c_M}^{(i)} \qquad \bigl( 1 \leq \tau \leq T, \quad 1 \leq a \leq N,\quad 1\leq i \leq k \bigr), 
\end{align}
with their capacity bounds given as
\begin{align}
q_m \leq c_m \leq Q_m \qquad \bigl(q_m, Q_m \in \mathbb Z,\quad 1\leq m \leq M \bigr). 
\end{align}
In many cases, we use a shorter notation, $c = (c_1,c_2,\cdots,c_M)$, as a vector of $M$-dimensional capacitated variables.%
\footnote{We also use the following ``array'' notation: 
\begin{align}
q \leq c \leq Q\qquad \Leftrightarrow \qquad q_m \leq c_m \leq Q_m \quad \bigl(1 \leq m \leq M \bigr). 
\end{align}}
The introduction of capacitated variables is understood as the direct product of {\em capacity-qubits} and the original VRP-qubits (See Fig.~\ref{capacity-qubit-figure}). 
The interpretation of the binary values, Eq.~\eq{eq-VRP-qubits-arrival}, should be now incorporated with the status of capacitated variables as follows: 
\begin{align}
x_{\tau,a|c}^{(i)} = 1\qquad  \bigl( \text{arriving with the status $c$ of capacitated variables} \bigr). 
\end{align}


Furthermore, to describe variation of capacitated variables, we introduce {\em variation matrices} 
$\bigl( B_{ab|m}^{(\tau)} \bigr)_{1 \leq a\neq b \leq N}$ for each time-interval $(\tau)$:%
\footnote{It is also convenient to use a vector notation of the collection of capacity-variation matrices: 
\begin{align}
B_{ab}^{(\tau)} \equiv \bigl( B_{ab|1}^{(\tau)}, B_{ab|2}^{(\tau)}, \cdots, B_{ab|M}^{(\tau)} \bigr). 
\end{align}
}
\begin{align}
B_{ab|m}^{(\tau)} =  &\biggl(  \parbox[c]{8cm}{variation of capacitated variable $c_m$ for traveling  
from a city $(b)$ (at time $\tau$) to a city $(a)$} \biggr) \in \mathbb Z. 
\end{align}
We can write the Hamiltonian as follows: 
\begin{align}
\mathcal H = &  \sum_{ 
\left\{
\begin{subarray}{c}
1\leq a \neq b \leq N \cr
1\leq \tau \leq T -1 \cr
1\leq i \leq k \cr
q \leq c \leq Q
\end{subarray}
\right\}
} \left(
\begin{array}{l}
\ds \vspace{0.2cm} \frac{d_{ab}^{(\tau)} - \mu}{\rho} \times x_{\tau + n_{ab}^{(\tau)}, a \big| c+B_{ab}^{(\tau)}}^{(i)} \, x_{\tau,b |c}^{(i)} \quad + \vspace{0.3cm}  \cr
\ds  \qquad + \sum_{q\leq c' (\neq c+B_{ab}^{(\tau)}) \leq Q}  \underbrace{\lambda \times x_{\tau + n_{ab}^{(\tau)},a| c'}^{(i)} \, x_{\tau,b|c}^{(i)}  }_{(\ast 4)}
+   \vspace{0.2cm} \cr
\ds \qquad + \sum_{
\begin{subarray}{c}
1\leq \delta \tau \leq n_{ab}^{(\tau)} - 1 \cr
q\leq c' \leq Q
\end{subarray}
} \underbrace{ \lambda \times x_{\tau + \delta \tau, a|c'}^{(i)} \, x_{\tau, b|c}^{(i)}  }_{(\ast 5)}
\end{array}
\right) + \nn\\
+& \lambda\times \left( 
\begin{array}{c} 
\ds \vspace{0.2cm}
\sum_{
\left\{
\begin{subarray}{c}
1\leq a \leq N \cr
1\leq \tau \leq T \cr
1\leq i \leq k  \cr
q\leq c'< c \leq Q
\end{subarray}
\right\}
} 
\underbrace{x_{\tau,a|c'}^{(i)} \, x_{\tau,a|c}^{(i)}}_{(\ast 6)} + 
\sum_{ 
\left\{
\begin{subarray}{c}
1\leq a < b \leq N \cr
1\leq \tau \leq T \cr
1\leq i \leq k \cr
q\leq c',  c \leq Q
\end{subarray}
\right\}
} 
x_{\tau,a|c'}^{(i)} \, x_{\tau,b|c}^{(i)}    + \cr
\ds 
+ \sum_{ 
\left\{
\begin{subarray}{c}
1\leq a \leq N \cr
1\leq \tau \neq \tau' \leq T \cr
1\leq i \leq j \leq k \cr
q\leq c',  c \leq Q
\end{subarray}
\right\}
} 
x_{\tau',a|c'}^{(i)} \, x_{\tau,a|c}^{(j)}    +
\sum_{ 
\left\{
\begin{subarray}{c}
1\leq a \leq N \cr
1\leq \tau \leq T \cr
1\leq i <  j \leq k \cr
q\leq c',  c \leq Q
\end{subarray}
\right\}
} 
x_{\tau,a|c'}^{(i)} \, x_{\tau,a|c}^{(j)}
\end{array}
\right). 
\label{Hamiltonian-2}
\end{align}
As is in Eq.~\eq{Hamiltonian-1}, the first line includes the interaction associated with consumption of time and variation of capacitated variables. In particular, 
\begin{itemize}
\item The second term $(\ast 4)$ in the first line ensures that transmission of capacitated variables in each travel satisfies the expected variation relation, 
\begin{align}
c \quad \mapsto \quad c' = c+ B_{ab}^{(\tau)}, 
\end{align}
\item The third term $(\ast 5)$ forbids early arrivals at a city $(a)$ from a city $(b)$ with any changes of capacitated variables, $c\to c'$. 
\end{itemize}
The second part represents the basic constraints of penalty terms: 
\begin{itemize}
\item [1. ] The newly introduced term is the first term $(\ast 6)$ and it represents that 
a vehicle $(i)$ arriving at a city $(a)$ in a time-interval $(\tau)$ cannot be assigned more than one capacitated-variable status, i.e.,~$c' \neq c$. 
\begin{align}
\lambda \times x_{\tau,a|c'}^{(i)} \, x_{\tau,a|c}^{(i)} \qquad \bigl( \forall a, \forall i, \forall \tau, \forall c' \neq \forall c \bigr). 
\end{align}
\item [2. ]  The other terms are essentially the same as in the previous section. 
\end{itemize}
As is in the previous section as well as Section \ref{section-CC-cond}, the parameters $(\mu, \rho, \lambda)$ should be chosen as the same standard values as stated in Eq.~\eq{Eq-mu-rho-parameter-choice}. 

A new feature obtained by our capacity formulation is that the variation matrices $B_{ab|m}^{(\tau)}$ can take any integer number. Therefore, the capacitated variables can increase and/or decrease (i.e.,~pickup and/or delivery), strictly satisfying the capacity bounds.

\section{State of vehicles \label{Section-state}}

In addition to the simple concept of capacity, the capacitated variables can also be interpreted as {\em the states of vehicles}. Through the use of ``state'', we can introduce various traveling rules depending on each state of vehicles. For the sake of simplicity, we shall demonstrate {\em two-state TSP/VRP} here. 

As a simple example of a two-state TSP/VRP, we shall consider the arrival-state and departure-state of the vehicles. In terms of qubits, the states of vehicle are now denoted by the hat ``$\widehat{\ \ }$'' above the binary variables: 
\begin{align}
\widehat x_{\tau, a}^{(i)}: & \quad \text{arrival-qubits},  & x_{\tau, a}^{(i)}: & \quad \text{departure-qubits}. 
\end{align}
A new ingredient of introducing two states is that we can now describe traveling/staying phases of vehicle: 
\begin{align}
1) \quad x \to \widehat x: & \quad \text{traveling phase},  &2) \quad 
\widehat x \to x: & \quad \text{staying phase}, 
\end{align}
and other transitions, say $x\to x$ and $\widehat x \to \widehat x$, should be forbidden. As an additional phase of motion is introduced, we can further add a new time-duration matrix $\widehat n_{a}^{(\tau)}$ to describe {\em how long the vehicle stays at the city $(a)$}: 
\begin{align}
 n_{ a}^{(\tau)} & = 
\biggl \lceil \frac{  \bigl(  \text{time spent for staying at a city $(a)$ 
before departure} \bigr)  }{\Delta t} \biggr \rceil \geq 1. 
\end{align}
Therefore, various traveling rules are additionally introduced as follows:%
\footnote{``Hermitian conjugate'' of the variables is defined as: $
\widehat {x_{\tau,a}^{(i)}} =\widehat  x_{\tau, a}^{(i)}$ and $\widehat {\widehat x_{\tau, a}^{(i)}} = x_{\tau, a}^{(i)}$. }
\begin{align}
\mathcal H & =  \sum_{ 
\left\{
\begin{subarray}{c}
1\leq a \neq b \leq N \cr
1\leq \tau \leq T -1 \cr
1\leq i \leq k
\end{subarray}
\right\}
}^{\text{``traveling''}}
\left( 
\begin{array}{c}
\ds  \vspace{0.1cm}
\frac{d_{ab}^{(\tau)} - \mu}{\rho} \times \widehat x_{\tau + n_{ab}^{(\tau)},  a}^{(i)} \, x_{\tau,b}^{(i)} 
+ \sum_{1\leq \delta \tau \leq n_{ab}^{(\tau)} - 1} \lambda \times \widehat x_{\tau + \delta \tau, a}^{(i)} \, x_{\tau, b}^{(i)}+ \cr 
\ds + \sum_{1 \leq \delta \tau \leq n_{ab}^{(\tau)} } \lambda \times x_{\tau + \delta \tau, a}^{(i)} x_{\tau, b}^{(i)} 
\end{array}
\right) + \nn\\
& +\sum_{ 
\left\{
\begin{subarray}{c}
1\leq a, b \leq N \cr
1\leq \tau \leq T -1 \cr
1\leq i \leq k
\end{subarray}
\right\}
}^{\text{``staying''}}
\left( 
\begin{array}{c}
\ds \vspace{0.1cm}
\delta_{ab} \biggl(
\frac{0 - \mu}{\rho} \times x_{\tau + n_a^{(\tau)}, a} \widehat x_{\tau, a}^{(i)}
+ 
\sum_{1\leq \delta \tau \leq n_{a}^{(\tau)} -1} \lambda\times x_{\tau + \delta \tau, a}^{(i)} \widehat  x_{\tau, a}^{(i)} \biggr) \cr
\ds +(1 - \delta_{ab}) \sum_{1 \leq \delta \tau \leq n_{b}^{(\tau)}} 
\lambda \times \widehat x_{\tau + \delta \tau, a}^{(i)} \widehat x_{\tau, b}^{(i)} 
\end{array}
\right)
+ \nn\\ 
& + \lambda\times 
\left( 
\begin{array}{l}
\ds \vspace{0.1cm}
\sum_{ 
\left\{
\begin{subarray}{c}
1\leq a \leq N \cr
1\leq \tau \leq T \cr
1\leq i \leq k 
\end{subarray}
\right\}
} 
\widehat x_{\tau, a}^{(i)} \, x_{\tau,a}^{(i)}    +
\sum_{ 
\left\{
\begin{subarray}{c}
1\leq a < b \leq N \cr
1\leq \tau \leq T \cr
1\leq i \leq k 
\end{subarray}
\right\}
} 
\Bigl( x_{\tau,a}^{(i)} \, x_{\tau,b}^{(i)}    + \widehat x_{\tau, a}^{(i)} \, x_{\tau,b}^{(i)} \Bigr) + \cr
\ds  \vspace{0.1cm} \qquad + \sum_{ 
\left\{
\begin{subarray}{c}
1\leq a \leq N \cr
1\leq \tau \neq \tau' \leq T \cr
1\leq i \leq j \leq k 
\end{subarray}
\right\}
} 
\Bigl( x_{\tau',a}^{(i)} \, x_{\tau,a}^{(j)}    + \widehat x_{\tau', a}^{(i)} \, x_{\tau,a}^{(j)}  \Bigr) + \cr
\ds \qquad  + \sum_{ 
\left\{
\begin{subarray}{c}
1\leq a \leq N \cr
1\leq \tau \leq T \cr
1\leq i <  j \leq k 
\end{subarray}
\right\}
} 
\Bigl( 
x_{\tau,a}^{(i)} \, x_{\tau,a}^{(j)} + 
\widehat x_{\tau, a}^{(i)} \, x_{\tau,a}^{(j)}
\Bigr) + \cr
 \qquad \qquad \qquad \qquad \qquad \qquad + \Bigl( \text{``hermitian conjugate''} \Bigr)
\end{array}
\right). \label{Hamiltonian-3}
\end{align}
Construction of the Hamiltonian is essentially the same as that of capacity (particularly the last line, which was obtained by re-interpreting capacitated variables $c$ of Section \ref{section-capacity-qubits} as states). A distinguishing point is that we can introduce different cost matrices $d_{ab}^{(\tau)}$ and duration matrices $n_{ab}^{(\tau)}$ depending on the state (i.e.,~phase) of each vehicle. 

This system (including ``states'' of vehicle) is referred to as {\em time-scheduled state-vehicle routing problem} or {\em TS-SVRP}. As a trivial extension, one can also consider multiple-state models and also the full system of {\em time-scheduled multiple-capacitated state-vehicle routing problem} or {\em TS-mCSVRP}.

\section{Practical constraints with time, state and capacity\label{Section-other}}

In practical applications for mobility service, there are various kinds of constraints that should be {\em simultaneously} implemented. The following list describes examples of such constraints, accepted by our TS-mCSVRP formulation: 
\begin{itemize}
\item [1) ] {\bf setting:}\quad  Consider an optimization problem for a delivery service. Cost of delivery is the total delivery time, and there are 50 customers, which are shared by five vehicles. 
\begin{itemize}
\item Five vehicles comprises three middle-size cars and two trucks. 
\item Delivery is served in eight hours (e.g.,~from 9:00 to 17:00). 
\item Delivery schedule is set by a twenty-minute unit. 
\item The cost of optimization would be the total consumption of delivery time. 
\end{itemize}
\item [2) ] {\bf time-variation:} \quad Traffic conditions can change depending on the delivery time. Therefore, the cost $d_{ab}^{(\tau)}$ depends on time $\tau$. Similarly, the cost of inbound and outbound routes is different in general, i.e.,~$d_{ab}^{(\tau)} \neq d_{ba}^{(\tau)}$. 
\item [3) ] {\bf priority for delivery:} \quad Some customers may request a priority for delivery, which can be achieved by adding some extra weight factor to the cost function, $d_{ab}^{(\tau)} \to \varphi* d_{ab}^{(\tau)}$\,  ($0<\varphi < 1$). 
\item [4) ] {\bf time-window and type-window:} \quad There are three kinds of constraints associated with {\em window}. These constraints can be realized by turning off the associated binary qubits (i.e.,~$x_{\tau, a|c}^{(i)} = 0$) by hand:
\begin{itemize}
\item Each customer has a request of {\em delivery time}, scheduled by twenty minutes. The delivery-time request can be multiple (i.e.,~disconnected) time-window for each customer. 
\item Some of the vehicles cannot deliver to some customers; e.g.~the roads are too narrow for trucks and some of the packages are too large or heavy for small cars to deliver. This requires {\em type-window} for delivery service. 
\item Each vehicle should serve within their own working hours. In particular, some of the drivers work in a short time. This requires {\em time-window for vehicles}. 
\end{itemize}
\end{itemize}
Notably, only the formulation of time-window has a discrepancy with our capacity and state description as it induces overtime traveling sometimes (See Section \ref{section-CC-cond}). Except for time-window, one can further implement the following two constraints: 
\begin{itemize}
\item [5) ] {\bf capacity-constraints:} \quad Capacity constraints are realized using capacity-qubits: 
\begin{itemize}
\item Each vehicle has its own volume and weight limitations for their capacity. 
\item For some vehicles that are used for both pickup and delivery, they need to be scheduled without over-capacity for volume and weight in delivering their service. 
\end{itemize}
\item [6) ] {\bf scheduling as state:} \quad Some detail about the schedule can be described using {\em state-description}. 
\begin{itemize}
\item After arriving at each customer's premise, drivers should spend twenty minutes on their customer services. 
\item Drivers can take a one-hour rest for lunch. This can be formulated using the time-dependence of duration matrices $n_{ab}^{(\tau)}$ or $n_{a}^{(\tau)}$. 
\end{itemize}
\end{itemize}

\section{Validity of the formulation with D-Wave 2000Q \label{section-CC-cond}}

Thus far, we have discussed the proposed new QUBO formulation of TSP/VRP. Conversely, it is also important to discuss the validity of the proposed formulation using a quantum annealing machine, D-Wave 2000Q.


\begin{figure}
\includegraphics[width=\textwidth]{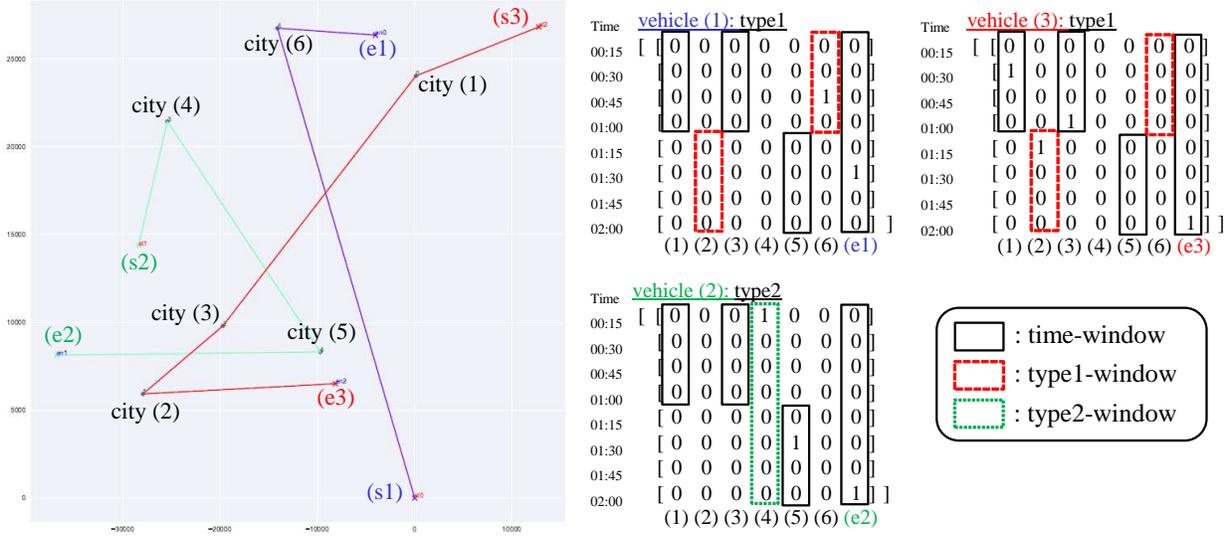}
\caption{Graphical view of the solution for a VRP-instance: the lowest energy state found with $\text{\tt num\_reads} = 10,000$ (i.e.,~time-to-soluition, ${\rm tts} \simeq 20\mu{\rm sec}\times 10,000 = 0.2 {\rm sec}$) on the D-Wave QPU, {\tt DWave2000Q\_2\_1}. Minor embedding is processed by the {\tt find\_embedding} utility provided by D-Wave System Inc., where {\tt minimize\_energy} is used for the broken chain treatment. Among the 10,000 samples, $80\%$ of the solutions are feasible solutions, which do not receive any penalty contributions. } \label{vrp-evaluate-figure}
\end{figure}


Fig.~\ref{vrp-evaluate-figure} shows a primitive instance of a delivery service (TS-VRP with windows) for six customers with three vehicles in two hours. The unit of time-scheduling is chosen to be fifteen minutes. Interestingly, a travel from the city (4) to the city (5) automatically chooses overtime traveling, because time-window occasionally prevents the shortest travel. 

Note that the parameters $(\rho, \mu, \lambda)$ are chosen as the standard value, Eq.~\eq{Eq-mu-rho-parameter-choice}, and therefore further optimization of the parameters would improve the performance. With taking into account this result, this instance shows that quantum annealing of our formulation works properly as far as small-size QUBO systems ($\simeq 83$ logical qubits for this instance) are considered.

\subsection{Negatively-shifted energy method and choice of baselines}
Further clarification is needed about our proposed method for the basic constraints (discussed around Eq.~\eq{Eq-mu-rho-parameter-choice}). As mentioned earlier, we did not apply the conventional method of ``square of linear-constraint functions,'' as in \cite{Lucas2014}. Instead, we apply {\em a negatively-shifted energy method}; therefore we will discuss how our proposed method can replace the conventional method. 

We consider a typical Hamiltonian, which can be generally expressed as follows: 
\begin{align}
\mathcal H = \sum_{A,B = 1}^N \underbrace{\bigl( C_{AB} - \mu \bigr)}_{\equiv \epsilon_{AB}}\times  x_A \, x_B - \xi \sum_{A=1}^N x_A+  \sum_{\vev{A,B} \in E_\text{frb.}} \lambda \times x_A \, x_B, 
\end{align}
where $C_{AB} \, (>0)$ is the general cost matrix and $E_{\text{fbd}}$ is a set of forbidden pairs (i.e.,~edges) of configurations. We usually choose $\xi=0$. 

In the conventional method (See \cite{Lucas2014}), the parameter $\xi$ inevitably exists and is strictly correlated with $\lambda \, (>0)$ as follows: 
\begin{align}
\xi = \lambda \quad (\text{TSP}),\qquad \xi \geq \frac{3}{2} \lambda \quad ( \text{VRP} ),  \label{EqnXiLambdaComv}
\end{align}
and $\mu = 0$. 
If one forgets about the correlation and chooses $\xi = 0$ and $\lambda>  0$ (and $\mu = 0$ in the conventional method), the ground state becomes trivial, $x_A = 0$ ($A=1,2,\cdots,N$). Therefore, the basic role of $\xi$-term is to enhance spontaneous popping-up of qubits: $x_A \to 1$ ($A = 1,2,\cdots,N$).

However, such effect can also be generated using {\em the overall negative shift of cost energy} obtained from $\mu$. This is possible since optimization only cares about the relative values of cost. Hence we can adjust $\mu$ such that 
\begin{align}
\epsilon_{AB} = C_{AB} - \mu < 0,  \label{negative-energy-eq}
\end{align}
for all the pairs $\left<A,B \right>$ focused.%
\footnote{This means that $\epsilon_{AB} \leq C_{\rm max}^{(\rm focused)} - \mu = 0  \leq C_{\rm max} - \mu$ for all the focused pairs $\vev{A,B}$. Therefore, the parameter $\mu$ plays a role of ``cut-off scale'' of cost energy.  } In the proposed method, we can set $\xi = 0$ or utilize it for other purposes. 

In a sequence of configurations generated using the negative energy, Eq.~\eq{negative-energy-eq}, there are forbidden configurations, which receive penalty caused by $\lambda$. As these configurations are forbidden, we impose that such configurations receive relatively positive energy. In our proposed VRP, this condition is given by 
\begin{align}
\epsilon_{AB} + \lambda > 0, \label{BaselineCondition-eq}
\end{align}
for all pairs $\left<A,B \right>$. 
We put a baseline (like a coastline) as energy $=0$ to separate feasible configurations (inside sea) from forbidden configurations (on the continent). This is referred to as the baseline condition. 

From these two conditions, we choose the maximum range of cost energy, 
\begin{align}
-\lambda = \epsilon_{\rm min } \leq \epsilon_{AB} \leq \epsilon_{\rm max} = 0, 
\end{align}
as the standard values. The larger range of cost causes larger resolution in quantum annealing machines. The solution of these conditions is given by Eq.~\eq{Eq-mu-rho-parameter-choice}. 

It is also possible to strategically choose higher baseline in Eq.~\eq{BaselineCondition-eq} and/or to select some focused energy range of Eq.~\eq{negative-energy-eq} as
\begin{align}
\epsilon_{AB} + \lambda + \delta > 0 \quad \Rightarrow \quad 
-\lambda -\delta = \epsilon_{\rm min } \leq \epsilon_{AB} \leq \epsilon_{\rm max}^{(\rm focused)} = 0, 
\end{align}
which may improve the performance. Despite increase in forbidden configurations, it can further improve the range of energy input. 
In contrast, too large $\delta(>0)$ reduces performance, because the probability of ground states evaporates into the forbidden configurations, which possess much larger entropy.%
\footnote{This phenomenon also occurs in the conventional method: if the parameter $\xi$ becomes too large, it induces a large number of forbidden configurations. Therefore, too large $\xi$ also reduces the performance of quantum annealing. }

\section{Conclusion and discussion \label{Section-conclusion}}

In this paper, we proposed a new QUBO formulation of CVRP with time, state, and capacity. Introduction of the strict concept of time allowed us to formulate various constraints associated with time, while the introduction of capacity-qubits allowed us to formulate pickup and delivery during each travel of the vehicles. Introduction of state allowed us to describe various traveling rules depending on the state of the vehicles. We evaluated the proposed QUBO formulation using a quantum annealing machine,  D-Wave 2000Q and the results show that our formulation properly works in small-size QUBO systems (less than $90$ logical qubits $\simeq$ $6 \sim 7$ customers), which can be directly embedded in the current D-Wave machines. 

For real-world applications of this formulation, on the other hand, at least more than $2000$ logical qubits ($\simeq$ more than $30$ customers for 20-minute scheduling of a half day) are required. In this sense, it is also interesting to evaluate our proposed TS-mCSVRP on digital Ising machines. It is also important to develop an efficient quantum/classical hybrid algorithm for our TS-mCSVRP, which should hasten the practical usage of our formulation. This point shall be reported in the future communication.

\vspace{1cm}
\noindent
{\bf \large Acknowledgment}
\vspace{0.2cm}

\noindent
This work is based on a joint project of DENSO Corporation, Toyota Tsusho Corporation and Toyota Tsusho Nexty Electronics (Thailand) Co., LTD. The authors greatly thank Shunsuke Takahashi, Masakazu Gomi and Toru Awashima for valuable discussions which make this work possible.


\begin{thebibliography}{99}

\bibitem{VRP-first-reference} 
  Dantzig, G.~B., Ramser,  J.~H.: 
  ``The Truck Dispatching Problem,''
  Management Science, Vol.~6, No.~1 (1959), pp.~80-91 
  [10.1287/mnsc.6.1.80]

\bibitem{Kadowaki-Nishimori1998}
  Kadowaki, T., Nishimori, H.:
  ``Quantum Annealing in the Transverse Ising Model,''
  Phys. Rev. E 58 (1998) pp.~5355-5363, 
  [10.1103/PhysRevE.58.5355]
  
\bibitem{DWaveCite}
  Johnson, M.~W., Amin, M.~H.~S., Gildert, S., Lanting, T., Hamze, F., Dickson, N., Harris, R., Berkley, A.~J., Johansson, J., Bunyk, P., Chapple, E.~M., Enderud, C., Hilton, J.~P., Karimi, K., Ladizinsky, E., Ladizinsky, N., Oh, T., Perminov, I., Rich, C., Thom, M.~C., Tolkacheva, E., Truncik, C.~J.~S., Uchaikin, S., Wang, J., Wilson, B., Rose, G.:
  ``Quantum annealing with manufactured spins,''
  Nature 473 (2011) pp.~194–198

\bibitem{Lucas2014}
  Lucas, A.:
  ``Ising formulations of many NP problems,''
  Front.\ Phys. Vol.~2 (2014) pp.~1-5, 
  [10.3389/fphy.2014.00005]

\bibitem{HopfieldTank1985}
  Hopfield, J.~J., Tank, D.~W.:
  ``Neural Computation of Decisions in Optimization Problems,''
  Biol.\ Cybern.\ 52 (1985), pp.~141-152, 
  [10.1007/BF00339943]
  
\bibitem{HopfieldTank1986}
  Hopfield, J.~J., Tank, D.~W.:
  ``Computing with neural circuits: a model,''
  Sience, New Series, Vol.~233, Issue 4764 (1986), pp.~625-633, 
  [10.1126/science.3755256]

\bibitem{MezardParisi1985} 
  M\`ezard, M., Parisi, G.: 
  ``Replicas and optimization,''
   J.\ Physique Lett.\ 46 (1985), pp.~771-778, 
   [10.1051/jphyslet:019850046017077100]
   
\bibitem{BrainPad2017}
Itoh, T., Ohta, M., Yamazaki, Y., Tanaka, S.:
``Quantum annealing for combinatorial optimization problems with multiple constraints,''
A poster in Adiabatic Quantum Computing Conference 2017 (AQC-17), 26 June 2017, Tokyo, Japan

\bibitem{QASAR2018}
  Feld, S., Gabor, T., as a jointwork with Volkswagen:
  A talk in Qubits Europe 2018 D-Wave Users Conference, 12 April 2018, Munich, Germany [available in {\tt https://www.dwavesys.com/sites/default/files/lmu-merged-published.pdf}]

  
 
\end{thebibliography}
\end{document}